\documentclass[12pt,letterpaper]{article}
\usepackage{t1enc} \usepackage{times}
\usepackage{amssymb}
\usepackage{amsmath}
\usepackage[mathscr]{euscript}
\usepackage{natbib}
\usepackage[totalwidth=437pt, totalheight=664pt]{geometry}
\DeclareMathAlphabet\scr{U}{scr}{m}{n}
\SetMathAlphabet\scr{bold}{U}{scr}{b}{n}
  \DeclareFontFamily{U}{scr}{\skewchar\font'177}%
  \DeclareFontShape{U}{scr}{m}{n}{<-6>rsfs5<6-8>rsfs7<8->rsfs10}{}%
  \DeclareFontShape{U}{scr}{b}{n}{<-6>rsfs5<6-8>rsfs7<8->rsfs10}{}%

\newcommand{\mal}{\stackrel{\mbox{\tiny$\bullet$}}{}}

\newcommand{\rr}{\mathbb R}  
\newcommand{\rp}{\mathbb R _+}

\newtheorem{satz}{Theorem}[section]

\newtheorem{lemma}[satz]{Lemma}
\newtheorem{cor}[satz]{Corollary}
\newtheorem{@definition}[satz]{Definition}

\newtheorem{@bsp}[satz]{Example}
\newenvironment{bsp}{\begin{@bsp}\rm}{\end{@bsp}}
\newtheorem{@assumption}[satz]{Assumption}
\newenvironment{assumption}{\begin{@assumption}\rm}{\end{@assumption}}
\newtheorem{@convention}[satz]{Convention}

\newtheorem{@remark}[satz]{Remark}
\newenvironment{bem}{\begin{@remark}\rm}{\end{@remark}}
\newenvironment{bemm}{\noindent {\bf Remarks.}}{}

\newcommand{\be}{\begin{enumerate}}
\newcommand{\ee}{\end{enumerate}}
\newcommand{\beq}{\begin{equation}}
\newcommand{\eeq}{\end{equation}}
\newcommand{\bea}{\begin{eqnarray}}
\newcommand{\eea}{\end{eqnarray}}
\newcommand{\beaa}{\begin{eqnarray*}}
\newcommand{\eeaa}{\end{eqnarray*}}

\newcommand{\bpf}{\noindent {\sc Proof.}\ }
\newcommand{\ep}{\hfill $\square $}

\newcommand{\E}{\scr E}

\newcommand{\B}{\scr B}

\newcommand{\til}{\widetilde}

\renewcommand{\epsilon}{\varepsilon}
\renewcommand{\theta}{\vartheta}
\renewcommand{\rho}{\varrho}

\begin{document}
\title{Utility maximization in models with\\ conditionally independent increments}
\author{Jan Kallsen\footnote{Mathematisches Seminar,
Christian-Albrechts-Universit\"at zu Kiel,
Christian-Albrechts-Platz 4,
D-24098 Kiel, Germany,
(e-mail: kallsen@math.uni-kiel.de).}  
\quad Johannes Muhle-Karbe\footnote{Fakult\"at f\"ur Mathematik,
Universit\"at Wien,
Nordbergstra\ss e 15,
A-1090 Wien, Austria,
(e-mail: johannes.muhle-karbe@univie.ac.at).} 
}
\date{}
\maketitle

\begin{abstract}
We consider the problem of maximizing expected utility from terminal wealth in models with stochastic factors. Using martingale methods and a conditioning argument, we determine the optimal strategy for power utility under the assumption that the increments of the asset price are independent conditionally on the factor process.\\

Key words: utility maximization, stochastic factors, conditionally independent increments, martingale method \\

\end{abstract}

\section{Introduction}\setcounter{equation}{0}
A classical problem in Mathematical Finance is to maximize expected utility from terminal wealth in a securities market (cf.\ \cite{karatzas.shreve.98, korn.97} for an overview). This is often called the \emph{Merton problem}, since it was first solved in a continuous-time setting by Merton \cite{merton.69, merton.71}. In particular, he explicitly determined the optimal strategy and the corresponding value function for power and exponential utility functions and asset prices modelled as geometric Brownian motions. 

Since then, these results have been extended to other models of various kinds. For L\'evy processes (cf.\ \cite{foldes.90,framstad.al.99,kallsen.goll.99b,benth.al.01}), the value function can still  be determined explicitly, whereas the optimal strategy is determined by the root of a real-valued function. For some affine stochastic volatility models (cf.\ \cite{kim.omberg.96, kraft.05, liu.07, kallsen.muhlekarbe.08}), the value function can also be computed in closed form by solving some ordinary differential equations, while the optimal strategy can again be characterized by the root of a real-valued function. 

For more general Markovian models, one is faced with more involved partial (integro-) differential equations that typically do not lead to explicit solutions and require a substantially more complicated verification procedure to ensure the optimality of a given candidate strategy (cf.\ e.g. \cite{zariphopoulou.01} for power and \cite{rheinlaender.steiger.06} for exponential utility). A notable exception is given by models where the stochastic volatility is independent of the other drivers of the asset price process. In this case, it has been shown that the optimal strategy is \emph{myopic}, i.e.\ only depends on the local dynamics of the asset price, cf.\ e.g.\ \cite{grandits.rheinlaender.02} for exponential and \cite{benth.al.03b, lindberg.06,delong.klueppelberg.08} for power utility. In particular, it can be computed without having to solve any differential equations.

In the present study, we establish that this generally holds for power utility, provided that the asset price has \emph{independent increments conditional on some arbitrary factor process}.  As in \cite{grandits.rheinlaender.02}, the key idea is to condition on this process, which essentially reduces the problem to studying processes with independent increments. This in turn can be done similarly as for L\'evy processes in \cite{kallsen.goll.99b}. In the following, we make this idea precise. We first introduce our setup of processes with conditionally increments and prove that general L\'evy-driven models fit into this framework if the stochastic factors are independent of the other sources of randomness. Subsequently, we then state and prove our main result in 
 3. Given condtionally independent increments of the asset price, it provides a pointwise characterization of the optimal strategy that closely resembles the well-known results for logarithmic utility (cf.\ e.g.\ \cite{kallsen.goll.01a}). Afterwards, we present some examples. In particular, we show how the present results can be used to study whether the maximal expected utility that can be achieved in affine models is finite. For the proof of our main result we utilize that exponentials of processes with conditionally independent increments are martingales if and only if they are $\sigma$-martingales. A proof of this result is provided in the appendix.

For stochastic background, notation and terminology we refer to the monograph of Jacod and Shiryaev \cite{js.03}. In particular, for a semimartingale $X$, we denote by $L(X)$ the set of $X$-integrable predictable processes and by $\varphi \mal X$ the stochastic integral of $\varphi \in L(X)$ with respect to $X$. Moreover, we write $\E(X)$ for the stochastic exponential of a semimartingale X. When dealing with stochastic processes, superscripts usually refer to coordinates of a vector rather than powers. By $I$ we denote the identity process, i.e.\ $I_t=t$.

\section{Setup}\setcounter{equation}{0}
Our mathematical framework for a frictionless market model is as follows. Fix a terminal time $T\in \rp$ and a filtered probability space $(\Omega, \scr{F},(\scr{F}_t)_{t \in [0,T]},P)$. We consider traded securities whose price processes are expressed in terms of multiples of a numeraire security. More specifically, these securities are modelled by their discounted price process $S$, which is assumed to be a semimartingale with values in $(0,\infty)^d$. We consider an investor whose preferences are modelled by a \emph{power utility function} $u(x)=x^{1-p}/(1-p)$ for some $p \in \rp \backslash\{0,1\}$ and who tries to maximize expected utility from terminal wealth. Her initial endowment is denoted by $v \in (0,\infty)$. \emph{Trading strategies} are described by $\rr^d$-valued predictable stochastic processes $\varphi= (\varphi^1, \ldots,\varphi^d) \in L(S)$, where $\varphi^i_t$ denotes the number of shares of security $i$ in the investor's portfolio at time $t$. A strategy $\varphi$ is called \emph{admissible} if its discounted \emph{wealth process} $V (\varphi) := v + \varphi \mal S$ is nonnegative (no negative wealth allowed). An admissible strategy is called \emph{optimal}, if it maximizes $\psi \mapsto E(u(V_T(\psi)))$ over all competing admissible strategies $\psi$.

We need the following very mild assumption. Since the asset price process is positive, it is equivalent to NFLVR by the fundamental theorem of asset pricing.

\begin{assumption}\label{a:NFLVR}
There exists an \emph{equivalent local martingale measure}, i.e. a probability measure $Q \sim P$ such that the $S$ is a local $Q$-martingale.
\end{assumption}

Since the asset price process $S$ is positive, Assumption \ref{a:NFLVR} and \cite[I.2.27]{js.03} imply that $S_{-}>0$ as well. By \cite[II.8.3]{js.03}, this means that there exists an $\rr^d$-valued semimartingale $X$ such that $S^i=S^i_0\scr{E}(X^i)$ for $i=1,\ldots,d$. We interpret $X$ as the returns that generate $S$ in a multiplicative way.  To solve the utility maximization problem, we make the following crucial structural assumptions on the \emph{return process} $X$.

\begin{assumption}\label{a:conditioning}
\begin{enumerate}
\item The semimartingale characteristics $(B^X,C^X,\nu^X)$ (cf.\ \cite{js.03}) of $X$ relative to some \emph{truncation function} such as $h(x)=x 1_{\{|x| \leq 1\}}$ can be written as
$$ B^X_t=\int_0^t b^X_s ds, \quad C^X_t=\int_0^t c^X_s ds, \quad \nu^X([0,t]\times G)=\int_0^t K^X_s(G) ds,$$
with predictable processes $b^X, c^X$ and a transition kernel $K^X$ from $(\Omega \times \rp,\scr{P})$ into $(\rr^d,\B^d)$. The triplet $(b^X,c^X,K^X)$ is called \emph{differential} or \emph{local characteristics} of $X$.
\item There is a process $y$ such that $X$ also is a semimartingale with local characteristics $(b^X,c^X,K^X)$ relative to the augmented filtration $\mathbf{G}:=(\scr{G}_t)_{t \in [0,T]}$ given by 
\begin{equation*}
\scr{G}_t:=\bigcap_{s>t} \sigma(\scr{F}_s \cup \sigma((y_r)_{r \in [0,T]})), \quad 0\leq t \leq T,
\end{equation*}
and such that  $b_t^X, c_t^X$ and $K_t^X(G)$ are $\scr{G}_0$-measurable for fixed $t \in [0,T]$ and $G \in \scr{B}^d$. By \cite[II.6.6]{js.03}, this means that $X$ has $\scr{G}_0$-\emph{conditionally independent increments}, i.e.\ it is a $\scr{G}_0$-\emph{PII}.
\end{enumerate}
\end{assumption}

\begin{bemm}
\begin{enumerate}
\item In the present general framework, modelling the stock prices as ordinary exponentials $S^i=S^i_0\exp(\til{X}^i)$, $i=1,\ldots,d$ for some semimartingale $\til{X}$ leads to the same class of models (cf.\ \cite[Propositions 2 and 3]{kallsen.04}).
\item The first part of  Assumption \ref{a:conditioning} essentially means that the asset price process has no fixed times of discontinuity. This condition is typically satisfied, e.g.\ for diffusions, L\'evy processes and affine processes. 
\item The second part of Assumption \ref{a:conditioning} is the crucial one. It means that the local dynamics of the asset price are determined by the evolution of the process $y$, which can therefore be interpreted as a \emph{stochastic factor process}. 
\end{enumerate}
\end{bemm}

In general, a semimartingale $X$ will not remain a semimartingale with respect to an enlarged filtration (cf.\ e.g.\ \cite[Chapter VI]{protter.04} and the references therein). Even if the semimartingale property is preserved, the characteristics generally do not remain unchanged. Nevertheless, we now show that some fairly general models satisfy this property if the factor process $y$ is independent of the other sources of randomness in the model.

\subsection*{Integrated L\'evy models}\label{ss:intLevy}
In this section, we assume that $X$ is modelled as 
\begin{equation}\label{DAktie}
X=y_{-} \mal B,
\end{equation}
for an $\rr^{d\times n}$-valued semimartingale $y$ and an independent $\rr^n$-valued L\'evy process $B$ with L\'evy triplet $(b^B,c^B,K^B)$. Furthermore, we suppose that the underlying filtration $\mathbf{F}$ is generated by $B$ and $y$ (or equivalently by $X$ and $y$ if $d=n$ and $y$ takes values in the invertible $\rr^{d \times d}$-matrices). The following result shows that Assumption \ref{a:conditioning} is satisfied in this case. 

\begin{lemma}\label{Characteristics}
Relative to both $\mathbf{F}$ and $\mathbf{G}$, $X$ is a semimartingale with $\scr{G}_0$-measurable local characteristics $(b^X,c^X,K^X)$ given by
\begin{gather*}
b^X=y_{-}b^B+\int(h(y_{-}x)-y_{-}h(x))K^B(dx), \quad c^X=y_{-}c^B y_{-}^{\top},\\
K^X(G)=\int 1_G(y_{-}x)K^B(dx) \quad \forall G \in \B^d \backslash\{0\}.
\end{gather*}
In particular, Assumption \ref{a:conditioning} is satisfied.
\end{lemma}

\bpf Since $B$ is independent of $y$ and $\mathbf{F}$ is generated by $y$ and $B$, it follows from \cite[Theorem 15.5]{bauer.02} that $B$ remains a L\'evy process (and in particular a semimartingale), if its natural filtration is replaced with either $\mathbf{F}$ or $\mathbf{G}$. Since the distribution of $B$ does not depend on the underlying filtration, we know from the L\'evy-Khintchine formula and \cite[II.4.19]{js.03} that $B$ admits the same local characteristics $(b^B,c^B,K^B)$ with respect to its natural filtration and both $\mathbf{F}$ and $\mathbf{G}$. Since $y_{-}$ is locally bounded and predictable relative to $\mathbf{F}$ and $\mathbf{G}$, the process $X$ is a semimartingale with respect to $\mathbf{F}$ and $\mathbf{G}$ by \cite[I.4.31]{js.03}. Its  characteristics can now be derived by applying \cite[Proposition 2]{kallsen.04}. The $\scr{G}_0$-measurability is obvious. \ep

\subsection*{Time-changed L\'evy models}\label{ss:tcLevy}
We now show that Assumption \ref{a:conditioning} also holds for time-changed L\'evy models. For Brownian motion, stochastic integration and time changes lead to essentially the same models by the Dambis-Dubins-Schwarz theorem. For general L\'evy processes with jumps, however, the two classes are quite different. More details concerning the theory of time changes can be found in \cite{jacod.79}, whereas their use in modelling is dealt with in \cite{carr.al.03,kallsen.04}. Here, we assume that the process $X$ is given by
\begin{equation}\label{DAktieTime}
X=\int_0^\cdot \mu(y_{s-})ds+B_{\int_0^\cdot y_{s-} ds},
\end{equation}
for a  $(0,\infty)$-valued semimartingale $y$, a measurable mapping $\mu: \rr \to \rr^d$ such that $\int_0^T|\mu(y_{s-})|ds < \infty$, $P$-a.s., and for an independent $\rr^d$-valued L\'evy process $B$ with L\'evy-Khintchine triplet $(b^B,c^B,K^B)$. Moreover, we suppose that the underlying filtration is generated by $X$ and $y$. We have the following analogue of Lemma \ref{Characteristics}.

\begin{lemma}\label{CharacteristicsTime}
Relative to both $\mathbf{F}$ and $\mathbf{G}$, $X$ is a semimartingale with $\scr{G}_0$-measurable local characteristics $(b^X,c^X,K^X)$ given by
\begin{equation*}
b^X=\mu(y_{-})+b^B y_{-},  \quad c^X=c^B y_{-}, \quad K^X(G)=K^B(G)y_{-} \quad \forall G \in \B^d.
\end{equation*}
In particular, Assumption \ref{a:conditioning} is satisfied.
\end{lemma}

\bpf Relative to $\mathbf{F}$, the assertion follows literally as in the proof of \cite[Proposition 4.3]{pauwels.07}. For the corresponding statement relative to the augmented filtration $\mathbf{G}$, let $Y=\int_0^\cdot y_s ds$ and $U_r:=\inf\{ q \in \rp: Y_q \geq r\}$.  Define the $\sigma$-fields
\begin{equation*}
\scr{H}_t:=\bigcap_{s>t} \sigma((B_q)_{q \in [0,s]}, (U_r)_{r \in \rp}).
\end{equation*}
Since $B$ is independent of $y$ and hence $Y$, it remains a L\'evy process relative to the filtration $\mathbf{H}:=(\scr{H}_t)_{t \in \rp}$. Its distribution does not depend on the underlying filtration, hence we know from the L\'evy-Khintchine formula and \cite[II.4.19]{js.03} that it is a semimartingale with local characteristics $(b^B,c^B,K^B)$ relative to $\mathbf{H}$. By \cite[Proposition 5]{kallsen.04} the time-changed process $(\til{B}_{\vartheta})_{\vartheta \in [0,T]}:=(B_{Y_{\vartheta}})_{\vartheta \in [0,T]}$ is a semimartingale on $[0,T]$ relative to the time-changed filtration $(\til{\scr{H}}_{\vartheta})_{\vartheta \in [0,T]}:=(\scr{H}_{Y_{\vartheta}})_{\vartheta \in [0,T]}$ with differential characteristics $(\til{b},\til{c},\til{F})$ given by
\begin{equation*}
\til{b}_{\vartheta}=b^B y_{\vartheta-}, \quad \til{c}_{\vartheta}=c^B y_{\vartheta-}, \quad \til{K}_{\vartheta}(G)=K^B(G)y_{\vartheta-} \quad  \forall G \in \B^d.
\end{equation*}
Furthermore, it follows as in the proof of \cite[Proposition 4.3]{pauwels.07} that  $\til{\scr{H}}_t=\scr{G}_t$ for all $t \in [0,T]$. The assertion now follows by applying \cite[Proposition 2 and 3]{kallsen.04} to compute the characteristics of $X$.\ep\\

\begin{bemm}
\begin{enumerate}
\item For the proof of Lemma \ref{CharacteristicsTime} we had to assume that the given filtration is generated by the process $(y,X)$ or equivalently $(Y,X)$. In reality, though, the integrated volatility $Y$ and the volatility $y$ typically cannot be observed directly. Therefore the canonical filtration of the return process $X$ would be a more natural choice. Fortunately, $Y$ and $y$ are typically adapted to the latter if $B$ is an infinite activity process (cf.\ e.g.\ \cite{winkel.01}).
\item A natural generalization of \eqref{DAktieTime} is given by models of the form
\begin{equation*}
X=\int_0^\cdot \mu(y^{(1)}_{s-},\ldots,y^{(n)}_{s-})ds+\sum_{i=1}^n B^{(i)}_{Y^{(i)}},
\end{equation*}
for $\mu: (0,\infty)^n \to \rr^d$, strictly positive semimartingales $y^{(i)}$, $Y^{(i)}=\int_0^{\cdot} y^{(i)}_s ds$ and independent L\'evy processes $B^{(i)}$, $i=1,\ldots,n$. If one allows for the use of the even larger filtration generated by all $y^{(i)}$, $B^{(i)}_{Y^{(i)}}$, $i=1,\ldots,n$ the proof of Lemma \ref{CharacteristicsTime} remains valid.  If $Y^{(i)}$ is interpreted as business time in some market $i$, this class of models allows assets to be influenced by the changing activity in different markets. 
\end{enumerate} 
\end{bemm}

\section{Optimal portfolios}\setcounter{equation}{0}

For asset prices with conditionally independent increments we can now characterize the solution to the Merton problem as follows.

\begin{satz}\label{t:conditioning}
Suppose Assumptions \ref{a:NFLVR}, \ref{a:conditioning} hold and assume there exists an $\rr^d$-valued process $\pi \in L(X)$ such that the following conditions are satisfied up to a $dP \otimes dt$-null set on $\Omega \times [0,T]$.
\begin{enumerate}
\item $K^X(\{x \in \rr^d: 1+\pi^{\top}x \leq 0\})=0$. 
\item $\int \left|x(1+\pi^{\top}x)^{-p}-h(x)\right|K^X(dx)<\infty$. 
\item For all $\eta \in \rr^d$ such that $K^X(\{x \in \rr^d: 1+\eta^{\top}x < 0\})=0$, we have
$$(\eta^{\top}-\pi^{\top})\left(b^X-p c^X \pi +\int \left(\frac{x}{(1+\pi^{\top}x)^p}-h(x)\right)K^X(dx)\right) \leq 0,$$
\item $\int_0^T |\alpha_s| ds<\infty$, where
\begin{align*}
\alpha:=(1-p)\pi^{\top} b^X&-\frac{p(1-p)}{2} \pi^{\top}c^X \pi\\
&+\int \left((1+\pi^{\top}x)^{1-p}-1-(1-p)\pi^{\top}h(x)\right)K^X(dx). \label{a:conditionalfinite}
\end{align*}
\end{enumerate}
Then there exists a $\scr{G}_0$-measurable process $\til{\pi}$ satisfying Conditions 1-4 such that the strategy $\varphi=(\varphi^1,\ldots,\varphi^d)$ defined as
\begin{equation}\label{Strategy}
\varphi^i_t:= \til{\pi}^i(t)\frac{v\scr{E}(\til{\pi} \mal X)_{t-}}{S^i_{t-}}, \quad i=1,\ldots,d,   \quad t \in [0,T],
\end{equation}
is optimal with value process $V(\varphi)=v \scr{E}(\til{\pi} \mal X)$. The corresponding maximal expected utility is given by 
$$E(u(V_T(\varphi)))=\frac{v^{1-p}}{1-p} E\left(\exp\left(\int_0^T \alpha_s ds\right)\right).$$
In particular, if $\pi$ is $\scr{G}_0$-measurable, it is possible to choose $\til{\pi}=\pi$.
\end{satz}

\bpf \emph{Step 1}: In view of Conditions 1-4, the measurable selection theorem \cite[Theorem 3]{sainte-beuve.74} and \cite[Proposition 1.1]{jacod.79} show the existence of $\til{\pi}$, since $(b^X,c^X,K^X)$ are $\scr{G}_0$-measurable by Assumption \ref{a:conditioning}. Hence we can assume without loss of generality that $\pi$ is $\scr{G}_0$-measurable, because we can otherwise pass to $\til{\pi}$ instead.

\emph{Step 2}: Since $\pi$ and hence $\varphi$ is $\mathbf{F}$-predictable by assumption and $\scr{F}_t \subset \scr{G}_t$ for all $t \in [0,T]$, it follows that $\varphi$ is $\mathbf{G}$-predictable as well. In view of Assumption \ref{a:conditioning}, the local characteristics of $X$ relative to $\mathbf{F}$ coincide with those relative to $\mathbf{G}$. Together with \cite[III.6.30]{js.03} this implies that we have $\pi \in L(X)$ and hence $\varphi \in L(S)$ w.r.t.\  $\mathbf{G}$, too.

\textit{Step 3}: The wealth process associated to $\varphi$ is given by
$$V(\varphi) = v+\varphi \mal S = v(1+\E(\pi \mal X)_{-} \mal (\pi \mal X))=v\E(\pi \mal X).$$ 
Since Condition 1 and \cite[I.4.61]{js.03} imply $V(\varphi)>0$, the strategy $\varphi$ is admissible.

\textit{Step 4}: Let $\psi$ be any admissible strategy. Together with Assumption \ref{a:NFLVR} and \cite[I.2.27]{js.03}, admissibility implies  $V(\psi)= 0$ on the predictable set $\{V_{-}(\psi) = 0\}$. Hence we can assume without loss of generality that $\psi=0$ on $\{V_{-}(\psi)=0\}$, because we can otherwise consider $\widetilde{\psi}:=1_{\{V_{-}(\psi)>0\}}\psi$ without changing the wealth process. Consequently, we can write $\psi=\eta V_{-}(\psi)$ for some $\mathbf{F}$-predictable process $\eta$. The admissibility of $\psi$ implies $\eta^{\top}_t \Delta X_t \geq -1$ which in turn yields
\begin{equation}\label{e:condadm}
K^X(\{x \in \rr^d: 1+\eta^{\top}_t x < 0\})=0
\end{equation}
outside some $dP \otimes dt$ null set. Moreover, it follows as above that $\psi \in L(S)$ w.r.t.\ $\mathbf{G}$ as well. Since $\int_0^T |\alpha_s|ds<\infty$ outside some $P$-null set by Condition 4, the process
\begin{equation*}
L_t:=\exp\left(\int_t^T \alpha_s ds\right)=L_0 \scr{E}\left(\int_{\cdot}^T \alpha_s ds\right)_t
\end{equation*}
is indistinguishable from a c\`adl\`ag process of finite variation and hence a $\mathbf{G}$-semimartingale, because $\pi$ and $(b^X,c^X,K^X)$ are $\scr{G}_0$-measurable.  The local $\mathbf{G}$-characteristics $(b,c,K)$ of $(L/L_0)V(\varphi)^{-p}V(\psi)$ can now be computed with \cite[Propositions 2 and 3]{kallsen.04}. In particular,  we get
$$K(G)=\int 1_G\left(\frac{L_{-}}{L_0}V_{-}(\varphi)^{-p}V_{-}(\psi)\left( \frac{1+\eta^{\top}x}{(1+\pi^{\top}x)^p}-1\right)\right)K^X(dx) ,$$
for all $G \in \B \backslash \{0\}$, which combined with Condition 2 yields
\begin{equation}\label{e:condsigma1}
\int_{\{|x|>1\}} |x| K(dx) <\infty
\end{equation}
outside some $dP \otimes dt$-null set. Moreover, insertion of the definition of $\alpha$ leads to
\begin{eqnarray}\label{e:condsigma3}
b&=&\int (h(x)-x)K(dx)\\
&&+\frac{L_{-}}{L_0}V_{-}(\varphi)^{-p}V_{-}(\psi)(\eta^{\top}-\pi^{\top})\left(b^X-pc^X \pi+\int \left(\frac{x}{(1+\pi^{\top}x)^p}-h(x)\right) K^X(dx)\right),\notag
\end{eqnarray}
and hence 
\begin{equation}\label{e:condsigma2}
b+\int (x-h(x))K(dx) \leq 0
\end{equation}
$dP \otimes dt$-almost everywhere on $\Omega \times [0,T]$ by \eqref{e:condadm} and Condition 3. In view of \eqref{e:condsigma1} and \eqref{e:condsigma2} the process $(L/L_0)V(\varphi)^{-p}V(\psi)$ is therefore a $\mathbf{G}$-supermartingale by \cite[Lemma A.2]{kallsen.kuehn.04} and \cite[Proposition 3.1]{kallsen.03}.

\emph{Step 5}: For $\psi=\varphi$, \eqref{e:condsigma1}, \eqref{e:condsigma3} and \cite[Lemma A.2]{kallsen.kuehn.04} show that $Z:=(L/L_0)(V(\varphi)/v)^{1-p}$ is a strictly positive $\sigma$-martingale. By \cite[Proposition 3]{kallsen.04} , $\log(Z)$ is a $\scr{G}_0$-PII, hence $Z$ and in turn $(L/L_0)V(\varphi)^{1-p}$ are $\mathbf{G}$-martingales by Lemma \ref{l:A1}.

\textit{Step 6}: Now we are ready to show that $\varphi$ is indeed optimal. Since $u$ is concave, we have
\begin{equation}\label{bla}
u(V_T(\psi)) \leq u(V_T(\varphi))+ u'(V_T(\varphi))(V_T(\psi)-V_T(\varphi))
\end{equation}
for any admissible $\psi$. This implies 
\begin{align*}
E(u(V_T(\psi))|\scr{G}_0) &\leq  E(u(V_T(\varphi))|\scr{G}_0) + L_0 E\left(\frac{L_T}{L_0} V_T(\varphi)^{-p}V_T(\psi)-\frac{L_T}{L_0} V_T(\varphi)^{1-p}\bigg|\scr{G}_0\right)\\
& \leq E(u(V_T(\varphi))|\scr{G}_0),
\end{align*}
because $(L/L_0)V(\varphi)^{-p}V(\psi)$ is a $\mathbf{G}$-supermartingale and $(L/L_0)V(\varphi)^{1-p}$ is a $\mathbf{G}$-martingale, both starting at $v^{1-p}$. Taking expectations, the optimality of $\varphi$ follows. Likewise, the $\mathbf{G}$-martingale property of $(L/L_0)V(\varphi)^{1-p}$ yields the maximal expected utility. \ep\\

\begin{bemm}
\begin{enumerate}
\item The first condition ensures that the wealth process of the optimal strategy is positive. It is satisfied automatically if the asset price process is continuous. In the presence of unbounded positive and negative jumps it rules out shortselling and leverage. The second condition is only needed to formulate the crucial Condition 3, which characterizes the optimal strategy. A sufficient condition for its validity is given by
$$b^X-pc^X\pi+\int \left(\frac{x}{(1+\pi^{\top}x)^p}-h(x)\right)K^X(dx)=0.$$
While one does not have to require NFLVR if this stronger condition holds as well, it is less general than Condition 3 in the presence of jumps, cf.\ \cite{hurd.04} for a related discussion.
\item The fourth condition ensures that the maximal \emph{conditional} expected utility is finite. However, the maximal \emph{unconditional} expected utility does not necessarily have to be finite if the utility function is unbounded for $p>1$.  By \cite[III.6.30]{js.03}, Condition 4 is automatically satisfied for $\pi \in L(X)$ if $X$ is continuous. 
\item Given the mild regularity condition 4, the optimal strategy at $t$ is completely described by the local characteristics at $t$, i.e.\ it is \emph{myopic}. This parallels well-known results for logarithmic utility (cf.\ e.g.\ \cite{ kallsen.goll.01a}). It is important to note, however, that whereas the optimal strategy is myopic in the general semimartingale case for logarithmic utility, this only holds for power utility if the return process $X$  has conditionally independent increments. Otherwise an additional non-myopic term appears, see e.g.\ \cite{kim.omberg.96,zariphopoulou.01,kraft.05}
\end{enumerate}
\end{bemm}

\section{Examples}\label{ss:condbsp}\setcounter{equation}{0}
We now consider some concrete models where the results of the previous section can be applied. For ease of notation, we consider only a single risky asset (i.e.\ $d=1$), but the extension to multivariate versions of the corresponding models is straightforward.

\subsection*{Generalized Black-Scholes models}\label{ss:genBS}\index{Black-Scholes model!generalized}\label{ss:genbs}
Let $B$ be a standard Brownian motion, $y$ an independent semimartingale and again denote by $I$ the identity process $I_t=t$. Consider measurable functions $\mu: \rr \to \rr$ and $\sigma: \rr \to (0,\infty)$ such that $\mu(y_{-}) \in L(I)$ and $\sigma(y_{-}) \in L(B)$ and suppose the discounted stock price $S$ is given by
$$ S=S_0\E(\mu(y_{-}) \mal I+\sigma(y_{-})\mal B).$$
For $X:=\mu(y_{-}) \mal I+\sigma(y_{-})\mal B$, \cite[II.4.19]{js.03} and \cite[Proposition 3]{kallsen.04} yield $b^X=\mu(y_{-})$ as well as $c^X=\sigma^2(y_{-})$ and $K^X=0$. In view of Lemma \ref{Characteristics}, Assumption \ref{a:conditioning} is satisfied. Define
$$ \pi:=\frac{\mu(y_{-})}{p\sigma^2(y_{-})}.$$
By Theorem \ref{t:conditioning} and the second remark succeeding it, the strategy $\varphi:=\pi v\E(\pi \mal X)/S$ is optimal provided that $\pi \in L(X)$. If $y_{-}$ is $E$-valued for some $E \subset \rr$, this holds true e.g.\ if the mapping $x \mapsto \mu(x)/\sigma^2(x)$ is bounded on compact subsets of $E$.

\begin{bem}
This generalizes results of \cite{delong.klueppelberg.08} by allowing for an arbitrary semimartingale factor process instead of a L\'evy-driven Ornstein-Uhlenbeck (henceforth OU) process. Notice however, that unlike \cite{delong.klueppelberg.08} we only consider utility from terminal wealth and do not obtain a solution to more general consumption problems.  Finiteness of the maximal expected utility is ensured in the case $p>1$ in our setup, which complements the results of \cite{delong.klueppelberg.08}. They consider the case $p \in (0,1)$ and prove that the maximal expected utility is finite subject to suitable linear growth conditions on the coefficient functions $\mu(\cdot)$ and $\sigma(\cdot)$ and an exponential moment condition on the driver of the OU process. 
\end{bem} 

\subsection*{Barndorff-Nielsen and Shephard (2001)}
If we set $\mu(x):=\kappa+\delta x$ for constants $\kappa, \delta \in \rr$, let $\sigma(x):=\sqrt{x}$ and choose an OU process
\begin{equation}\label{e:OU}
dy_t =-\lambda y_{t-}+dZ_{t}, \quad y_0 \in (0,\infty),
\end{equation}
for a constant $\lambda > 0$ and some subordinator $Z$ in the generalized Black-Scholes model above, we obtain the model of Barndorff-Nielsen and Shephard \cite{barndorff.shephard.01}. Since $y_t \geq y_0 e^{-\lambda T}>0$ in this case, 
\begin{equation*}
\pi:=\frac{\mu(y_{-})}{p\sigma^2(y_ {-})}=\frac{\kappa}{py_{-}}+\frac{\delta}{p}
\end{equation*}
is bounded and hence belongs to $L(X)$. Consequently, $\varphi_t=\pi V(\varphi)/S$ is optimal.

\begin{bem}
This recovers the optimal strategy obtained by \cite{benth.al.03b}. Similarly as in \cite{delong.klueppelberg.08}, \cite{benth.al.03b} considers the case $p \in (0,1)$ and proves that the maximal expected utility is finite subject to an exponential moment condition on the L\'evy measure $K^Z$ of $Z$. Our results complement this by ascertaining that the same strategy is always optimal (with not necessarily finite expected utility), as well as optimal with finite  expected utility in the case $p>1$. 
\end{bem}

\subsection*{Carr et.\ al (2003)}
In this section we turn to the time-changed L\'evy models proposed by \cite{carr.al.03}, i.e.\ we let
\begin{equation}\label{e:tcaffin}
X_t=\mu t+B_{\int_0^t y_s ds}, \quad \mu \in \rr,
\end{equation}
for a L\'evy process $B$ with L\'evy-Khintchine triplet $(b^B,c^B,K^B)$ and an independent OU process $y$ given by \eqref{e:OU}. By Lemma \ref{CharacteristicsTime}, Assumption \ref{a:conditioning} holds. Hence we obtain 

\begin{cor}\label{c:CarrOpt}
Suppose $B$ has both positive and negative jumps and assume there exists a process $\pi$ such that the following conditions are satisfied.
\begin{enumerate}
\item $K^B(\{x \in \rr^d: 1+\pi x \leq 0\})=0$. 
\item $\int_0^T\left(\int \left|x(1+\pi x)^{-p}-h(x)\right|K^B(dx)\right)dt<\infty$. 
\item For all $\eta \in \rr^d$ such that $K^B(\{x \in \rr^d: 1+\eta x < 0\}=0$, we have
$$(\eta-\pi)\left(\left(\frac{\mu}{y_{-}}+b^B\right)-p c^B \pi +\int \left(\frac{x}{(1+\pi x)^p}-h(x)\right)K^B(dx)\right) \leq 0.$$
\end{enumerate}
Then  there exists a $\scr{G}_0$-measurable process $\widetilde{\pi} \in L(X)$ satisfying Conditions 1-3 such that $\varphi=\widetilde{\pi} v \E(\pi \mal X)_{-}/S_{-}$ is optimal. 
\end{cor}

\bpf  Since $B$ has both positive and negative jumps, the model satisfies Assumption \ref{a:NFLVR} by \cite[Lemma 4.42]{muhlekarbe.09}. Moreover, $\pi$ is bounded by Condition 1. Hence it belongs to $L(X)$ and Condition 2 implies that Condition 4 of Theorem \ref{t:conditioning} is also satisfied. By Lemma \ref{CharacteristicsTime}, Conditions 1-3 imply Conditions 1-3 of Theorem \ref{t:conditioning}. Consequently, the assertion immediately follows from Theorem \ref{t:conditioning} .\ep\\

For $\mu=0$ one recovers \cite[Theorem 3.4]{kallsen.muhlekarbe.08}, where the optimal fraction $\pi$ of wealth invested into stocks can be chosen to be deterministic. For $\mu \neq 0$, the optimal fraction depends on the current level of the activity process $y$. As for the generalized Black-Scholes models above, it is important to emphasize that the optimal strategy $\varphi$ is only ensured to lead to finite expected utility in the case $p>1$. However, the results provided here allow us to complete the study of the case $p \in (0,1)$ for $\mu=0$ started in \cite{kallsen.muhlekarbe.08}. Using Corollary \ref{c:CarrOpt}, we can now show that if there exists $\pi \in \rr$ satisfying Conditions 1-3, the exponential moment condition in \cite[Theorem 3.4]{kallsen.muhlekarbe.08} is necessary and sufficient for the maximal expected utility to be finite. The key observation is that the random variable $\int_0^T \alpha_s ds$ from Theorem \ref{t:conditioning} turns out to be infinitely divisible for $\mu=0$.

\begin{cor}\label{cor:utility}
Let $\mu=0$ and suppose there exists $\pi \in \rr$ satisfying the conditions of Corollary \ref{c:CarrOpt}. Then the maximal expected utility corresponding to the optimal strategy $\varphi:=\pi v \scr{E}(\pi X)_{-}/S_{-}$ is always finite for $p>1$, whereas for $p \in (0,1)$ it is finite if and only if
\begin{equation}\label{e:neccsuff}
\int_0^T \int_1^\infty \exp\left(\frac{e^{-\lambda t}-1}{\lambda}Cz\right) K^Z(dz) dt <\infty
\end{equation}
where
\begin{align*}
C:=(p-1)b^B\pi+\frac{p(1-p)}{2}c^B \pi^2-\int \left((1+\pi x)^{1-p}-1-(1-p)\pi h(x)\right) K^B(dx).
\end{align*}
If the maximal expected utility is finite, it is given by 
\begin{align*}
E(u(V_T(\varphi)))&\\
=\frac{v^{1-p}}{1-p}&\exp\left(\int_0^T \left(b^Z \widetilde{\alpha}(s) +\int (e^{\widetilde{\alpha}(s)z}-1-\widetilde{\alpha}(s)h(z))K^Z(dz)\right)ds+\widetilde{\alpha}(0)y_0\right),
\end{align*}
for $\widetilde{\alpha}(t)= C (e^{-\lambda(T-t)}-1)/\lambda$.
\end{cor}

\bpf  After inserting the characteristics of $X$, Theorem \ref{t:conditioning} shows that the maximal expected utility is given by
\begin{equation}\label{e:condmaxu}
E(u(V_T(\varphi)))=\frac{v^{1-p}}{1-p}E\left(\exp\left(-C\int_0^T y_{t}dt\right)\right).
\end{equation}
The process $(y,\int_0^\cdot y_s ds)$ is an affine semimartingale by \cite[Proposition 2]{kallsen.04}, hence \cite[Corollary 3.2]{kallsen.04} implies that the characteristic function of the random variable $\int_0^T y_s ds$ is given by 
$$ E\left( \exp\left(iu \int_0^T y_s ds\right)\right)= \exp\left(i b u + \int \left(e^{i u x} -1-i u h(x)\right)K(dx)\right), \quad \forall u \in \rr,$$
with
$$K(G) 	:= \int_0^T \int 1_G\left(\frac{1-e^{-\lambda t}}{\lambda}z\right) K^Z(dz) dt, \quad \forall G \in \scr{B}$$
and
\begin{align*}
b			:=  b^Z & \left(\frac{e^{-\lambda T}-1 +\lambda T}{\lambda^2}\right)+y_0\left(\frac{1-e^{-\lambda T}}{\lambda}\right)\\
 &+ \int_0^T \int \left(h\left(\frac{1-e^{-\lambda t}}{\lambda}z\right)-\frac{1-e^{-\lambda t}}{\lambda} h(z)\right) K^Z(dz) dt.
\end{align*}
Since $K^Z$ is a L\'evy measure, i.e.\ satisfies $K^Z(\{0\})=0$ and integrates $1 \wedge |x|^2$, one easily verifies that $b$ is finite and $K$ is a L\'evy measure, too. By the L\'evy-Khintchine formula, the distribution of $\int_0^T y_s ds$ is therefore infinitely divisible. Consequently \eqref{e:condmaxu} and \cite[Corollary 11.6 and Theorem 25.17]{sato.99} yield that $E(u(V_T(\varphi)))$ is finite if and only if
$$ \int_{\{|x|>1\}} e^{-Cx} K(dx) = \int_0^T \int_{\{|(1-e^{-\lambda t})z/\lambda| >1\}}\exp\left(\frac{e^{-\lambda t}-1}{\lambda}Cz\right) K^Z(dz) dt$$
is finite. Since $\lambda>0$ and the L\'evy measure $K^Z$ of the subordinator $Z$ is concentrated on $\rp$ by \cite[21.5]{sato.99}, the assertion follows. For $p>1$, Condition 3 of Corollary \ref{c:CarrOpt} and the Bernoulli inequality show that $C$ is positive. Consequently, \eqref{e:neccsuff} is always satisfied.  \ep\\

Since the exponential moment condition in Corollary \ref{cor:utility} depends on the time horizon, it is potentially only satisfied if $T$ is sufficiently small. This resembles the situation in the Heston model, where the maximal expected utility can be infinite for some parameters and sufficiently large $T$, if $p \in (0,1)$ (cf.\ \cite{kallsen.muhlekarbe.08}). However, a qualitatively different phenomenon arises here. Whereas expected utility can only tend to infinity in a continuously in the Heston model, it can suddenly jump to infinity here. This means that the utility maximization problem is not stable with respect to the time horizon in this case.

\begin{bsp} \textbf{(Sudden explosion of maximal expected utility)}\label{bsp:explosion}
In the setup of Corollary \ref{cor:utility} consider $p \in (0,1)$, $K^B=0$, $b^B \neq 0$, $c^B=1$ and hence $C=(b^B)^2(p-1)/2p<0$. Define the L\'evy measure 
$$ K^Z(dz):= 1_{(1,\infty)}(z)\exp\left(\frac{C}{2\lambda}z\right)\frac{dz}{z^2},$$
and let $b^Z=0$ relative to the truncation function $h(z)=0$ on $\rr$. Setting $T_{\infty}:=\log(2)/\lambda$, we obtain 
\begin{equation*}
\int_1^{\infty} \exp\left(\frac{e^{-\lambda t}-1}{\lambda} C z\right) K^Z(dz) \begin{cases} \leq 1, & \mbox{for } t\leq T_{\infty}, \\ =\infty, &\mbox{for } t > T_{\infty}. \end{cases}
\end{equation*}
Consequently, by Corollary \ref{cor:utility}, the maximal expected utility that can be obtained by trading on $[0,T]$ is finite for $T \leq T_{\infty}$ and satisfies
$$ E(u(V_T(\varphi)))\leq \frac{v^{1-p}}{1-p}\exp(\log(2)/\lambda+|C/2\lambda|y_0)<\infty.$$
Hence the maximal expected utility is actually bounded from above for $T \leq T_{\infty}$. For $T>T_{\infty}$, however, is is infinite by Corollary \ref{cor:utility}. 
\end{bsp}

Since $u(V_T(\varphi))=V_T(\varphi)^{1-p}/(1-p)$ is an exponentially affine process for $\mu=0$, the finiteness of the maximal expected utility is intimately linked to moment explosions of affine processes, cf.\ \cite{friz.kellerressel.09} and the references therein for more details.

\begin{appendix}

\section{Appendix}\setcounter{equation}{0}
In the proof of Theorem \ref{t:conditioning} we used that exponentials of processes with conditionally independent increments are martingales if and only if they are $\sigma$-martingales. In this appendix, we give a proof of this result.

\begin{lemma}\label{l:A1}
Let $X$ be an $\rr$-valued process with $X_0=0$ and conditionally independent increments relative to some $\sigma$-field $\scr{H}$. If $X$ admits local characteristics $(b,c,K)$ with respect to some truncation function $h$, the following are equivalent.
\begin{enumerate}
\item $\exp(X)$ is a martingale on $[0,T]$.\label{i:1}
\item $\exp(X)$ is a local martingale on $[0,T]$.\label{i:2}
\item $\exp(X)$ is a $\sigma$-martingale on $[0,T]$.\label{i:3}
\item Up to a $dP \otimes dt$-null set, we have $\int_{\{x>1\}} e^x K(dx)<\infty$ and  
\begin{equation}\label{e:eqdrift}
b+\frac{c}{2}+\int (e^{x}-1-h(x))K(dx)=0.
\end{equation}
\label{i:4}

\end{enumerate}
\end{lemma}

\bpf The implications \ref{i:1} $\Rightarrow$ \ref{i:2} $\Rightarrow$ \ref{i:3} follow from \cite[Lemma 3.1]{kallsen.03}. Moreover, \cite[Lemma 3.1]{kallsen.03} and \cite[Proposition 3]{kallsen.04} yield \ref{i:3} $\Leftrightarrow$ \ref{i:4}. Consequently, it remains to show \ref{i:4} $\Rightarrow$  \ref{i:1}. 

By \cite[Proposition 3.1]{kallsen.03}, the $\sigma$-martingale $\exp(X)$ is a supermartingale. Therefore it suffices to show $E(\exp(X_T))=1$. In view of \cite[Satz 44.3]{bauer.02} a regular version $R(\omega,dx)$ of the conditional distribution of $X_T$ w.r.t.\ $\scr{H}$ exists. From \cite[\S 44]{bauer.02} and \cite[II.6.6]{js.03} we get
\begin{align*}
&\int e^{iux} R(\omega,dx) \\
&\quad= E(\exp(iuX_T) | \scr{H})(\omega)\\   
&\quad= \exp\left(iuB_T(\omega)-\frac{1}{2}u C_T(\omega)u+\int_{[0,T] \times \rr^d} (e^{iux}-1-iuh(x))\nu(\omega,dt,dx)\right),
\end{align*}
where $B=b \mal I$, $C=c \mal I$ and $\nu=K \otimes I$ denote the semimartingale characteristics of $X$. By the L\'evy-Khintchine formula \cite[Theorem 8.1]{sato.99}, $R(\omega,\cdot)$ is therefore a.s.\ infinitely divisible. Since any supermartingale is a special semimartingale by \cite[Proposition 2.18]{jacod.79}, it follows from \cite[Corollary 3.1]{kallsen.03} that $\exp(X^i)$ is a local martingale. Hence
\begin{equation}\label{e:eqint}
\int_{[0,T] \times \{x>1\}} e^{x} \nu(dt,dx)<\infty, \quad \mbox{$P$-a.s.}
\end{equation}
by \cite[Proposition 3]{kallsen.04} and \cite[Lemma 3.1]{kallsen.03}. By \cite[Corollary 11.6 and Theorem 25.17]{sato.99}, \eqref{e:eqdrift} and \eqref{e:eqint} show that $\int e^{x} R(\omega,dx)=1$, $P$-a.s.\  and hence
$$E(\exp(X_T))=\int \int e^{x}R(\omega,dx) P(d\omega) =1.$$
This proves the assertion.	

\end{appendix}

\end{document}